\documentclass[%
 reprint,
superscriptaddress,
 amsmath,amssymb,
 aps,
]{revtex4-2}

\usepackage{graphicx}
\usepackage{dcolumn}
\usepackage{bm}
\usepackage[table]{xcolor} 
\renewcommand\vec[1]{\ensuremath\boldsymbol{#1}} 

\begin{document}

\title{Effects of nucleation at a first-order transition between two superconducting phases: Application to CeRh$_2$As$_2$}

\author{Andr\' as L. Szab\' o}
\affiliation{Institute for Theoretical Physics, ETH Zurich, 8093 Zurich, Switzerland}

\author{Mark H. Fischer}
\affiliation{Department of Physics, University of Zurich, 8057 Zurich, Switzerland}
\author{Manfred Sigrist}
\affiliation{Institute for Theoretical Physics, ETH Zurich, 8093 Zurich, Switzerland}

\date{\today}

\begin{abstract}

Recent experiments observed a phase transition within the superconducting regime of the heavy-fermion system CeRh$_2$As$_2$ when subjected to a $c$-axis magnetic field.
This phase transition has been interpreted as a parity switching from even to odd parity as the field is increased, and is believed to be of first order.
If correct, this scenario provides a unique opportunity to study the phenomenon of local nucleation around inhomogeneities in a superconducting context.
Here, we study such nucleation in the form of sharp domain walls emerging on a background of spatially varying material properties and hence, critical magnetic field.
To this end, we construct a spatially inhomogeneous Ginzburg-Landau functional and apply numerical minimization to demonstrate the existence of localized domain wall solutions and study their physical properties.
Furthermore, we propose ultrasound attenuation as an experimental bulk probe of domain wall physics in the system.
In particular, we predict the appearance of an absorption peak due to domain wall percolation upon tuning the magnetic field across the first-order transition line.
We argue that the temperature dependence of this peak could help identify the nature of the phase transition.
\end{abstract}

\maketitle


\emph{Introduction.} 
First-order phase transitions are ubiquitous in nature, with the most palpable everyday example being that of the liquid-gas transition of water. 
Increasing the vapor pressure in an air-water vapor mixture eventually triggers condensation of liquid water, seeded at inhomogeneities, such as dirt, in the medium.
Such local nucleation leads to the spatial coexistence of different phases and constitutes one of the hallmarks of first-order phase transitions.

In spite of the vast tableau of unconventional superconductors, to this day only a few compounds are known to feature multiple superconducting phases at ambient pressure.
For instance, UPt$_3$ exhibits three distinct pairing phases and an associated tetra-critical point in its magnetic field-temperature ($H$-$T$) phase diagram, albeit separated by second-order phase transition lines~\cite{joynt2002superconducting}. 
In contrast, the recently discovered CeRh$_2$As$_2$ is believed to undergo a first-order phase transition between two distinct superconducting phases upon increasing a $c$-axis magnetic field at sufficiently low temperature~\cite{khim2021field,ogata2023parity}. This feature provides an intriguing opportunity to study the effects of coexistence of the two phases for slightly inhomogeneous samples. 

\begin{figure}[t]
    \centering
    \includegraphics[width=8.3cm]{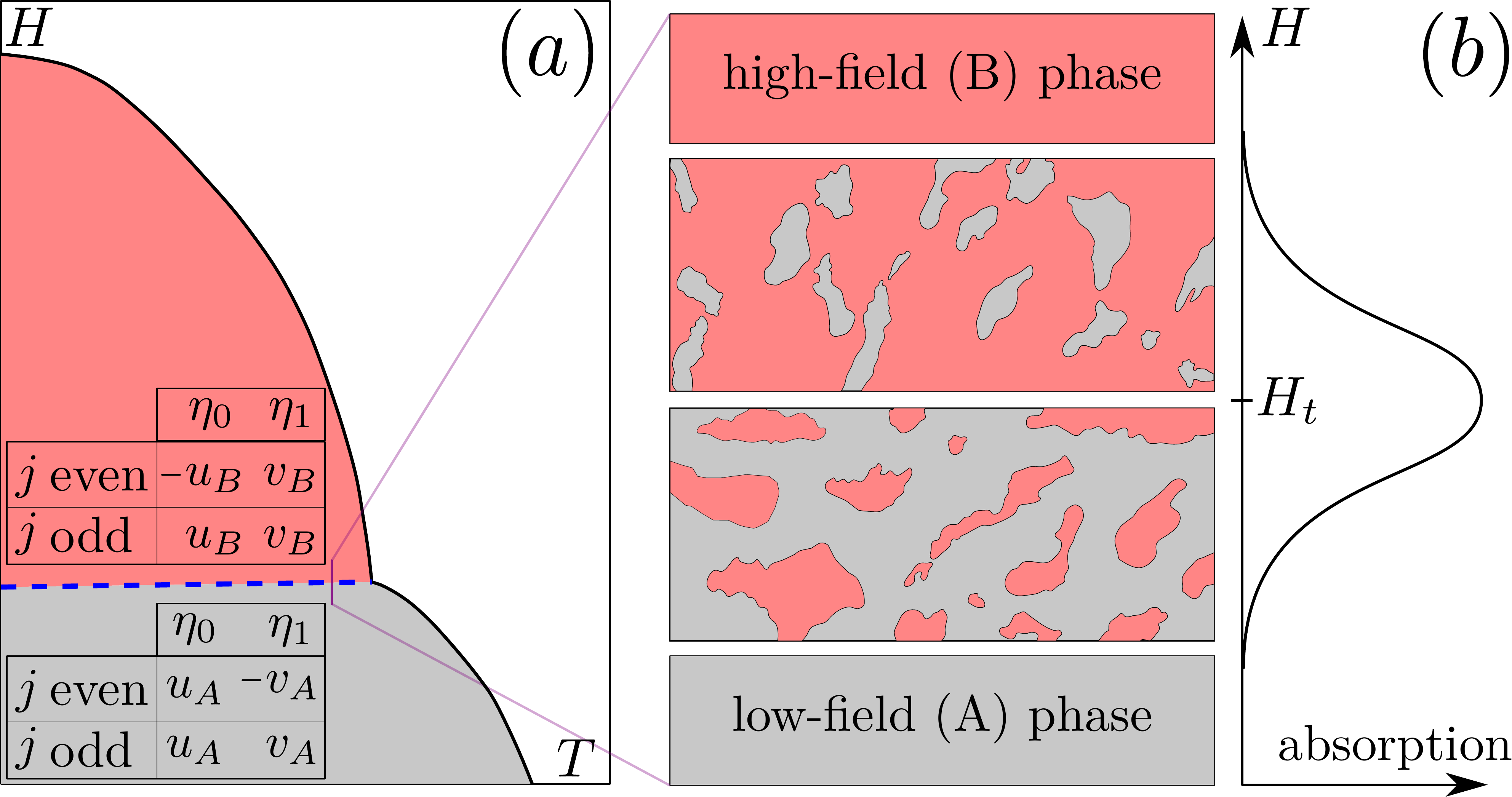}
    \caption{(a) Schematic phase diagram showing the even-parity low-field A (gray) and odd-parity high-field B (red) superconducting phases separated by a first-order internal phase transition (dashed blue) line at $H=H_t$. The inset shows the configuration of the order parameters $\eta_\mu^j$ in layer $j$ with $\mu=0,(1)$ for even (odd) parity. Here, $u_A>v_A>0$ and $v_B>u_B>0$. (b) Schematic depiction of nucleation. Deep in the A or B phase the system is devoid of domain walls. At $H\lesssim H_t$ ($H\gtrsim H_t$) the landscape is dominated by the A (B) phase, with bubbles of B (A). Maximum domain wall proliferation is expected at $H\approx H_t$.}
    \label{fig:phasediag}
\end{figure}

The heavy-fermion superconductor CeRh$_2$As$_2$ has been subject to intense experimental~\cite{kimura2021optical,landaeta2022field,kitagawa2022two,mishra2022anisotropic,onishi2022low,siddiquee2022tuning,semeniuk2023superconductivity,hafner2022possible,kibune2022observation,ogata2023parity} and theoretical~\cite{machida2022violation,hazra2022triplet,schertenleib2021unusual,nogaki2022even,nogaki2021topological,mockli2021two,ptok2021electronic,mockli2021superconductivity,cavanagh2022nonsymmorphic} investigation in recent years.
The most striking features of 
the $H$-$T$ phase diagram have been attributed to the staggered non-centrosymmetricity of this compound, where two inequivalent Ce sites (constituting a sublattice degree of freedom) experience different crystal fields from the surrounding Rh and As cages, giving rise to antisymmetric Rashba spin-orbit coupling (ASOC)~\cite{yoshida2015topological,cavanagh2022nonsymmorphic}.
While inversion symmetry is locally broken at each Ce atom, the point group of the material is $D_{4h}$, with an inversion center sitting between Ce sites.
ASOC induced by local non-centrosymmetricity and the resulting local parity mixing are believed to lend this system a surprising robustness against $c$-axis magnetic fields, with an extrapolated upper critical field of 14 T despite the low onset temperature of $T_{\rm c} \approx 0.26$~K to superconductivity~\cite{sigrist2014superconductors,skurativska2021spin}.
The upper critical magnetic field in this case rises in two stages when the temperature is lowered with a kink feature at $ T \approx 0.6 T_{\rm c} $ and $ H \approx 4$~T, see Fig.~\ref{fig:phasediag}(a).
This anomaly is associated with the first-order phase boundary line, which separates a low- and a high-field phase and is nearly temperature independent~\cite{khim2021field}.

In this work, we investigate consequences of material inhomogeneity on the properties of superconductivity in the vicinity of the first-order phase boundary in CeRh$_2$As$_2$.
Inhomogeneities give rise to nucleation when the phase boundary is approached upon tuning the magnetic field, with parts of the sample in the low- and others in the high-field phase, A and B, respectively, see Fig.~\ref{fig:phasediag}(b).
In the following, we use the phenomenological Ginzburg-Landau (GL) framework introduced earlier~\cite{schertenleib2021unusual} to study the formation and phenomenology of domain walls (DW).
As the balance of coexisting regions of A and B can be affected by pressure or strain, we study ultrasound coupling to the DWs, which are shaken and thus provide non-resonant attenuation upon their proliferation in the vicinity of the phase transition.


\emph{Model.} The unit cell of CeRh$_2$As$_2$ is spanned by two inequivalent layers (or sublattices) stacked in the crystallographic $c$ direction~\footnote{We use in the following a coordinate system with $x$ and $y$ the in-plane coordinates} with alternating Rashba-like ASOC.
A minimal GL model that reproduces the experimentally observed phase diagram is based on local parity-mixed $(s+p)$-wave pairing and associates the first-order phase boundary with a global parity-switching transition ~\cite{fischer2022superconductivity,schertenleib2021unusual}.
We introduce two layer-dependent order parameter (OP) components with layer index $j$, namely $\eta^j_0$, an $A_{1g}$ spin-singlet $s$-wave OP, and $\vec{d}^j_1=\eta^j_1 [k_y, -k_x, 0]$, the $d$-vector of the odd-parity component, transforming under the $A_{2u}$ representation of the point group $D_{4h}$.
The pertinent GL free energy density as an expansion in $\eta^j_0$ and $\eta^j_1$ is written as 
\allowdisplaybreaks
\begin{align}
    f &=\sum_{j} \sum_{\mu=0,1}  f_\mu^j + f_\epsilon + f_J + f_H, \text{ with} \nonumber \\
    f_{\mu}^j &= a_{\mu} |\eta_\mu^j|^2 + b_{\mu} |\eta_\mu^j|^4 + K_\mu |\vec{D}_\parallel \eta_\mu^j|^2, \nonumber \\
    f_\epsilon &= \sum_{j} (-1)^j \epsilon ( \eta_0^{j \ast} \eta_1^j + \eta_0^j \eta_1^{j \ast} ), \nonumber \\
    f_J &= J \sum_j \sum_{\mu} |\eta_\mu^{j+1} - \eta_\mu^{j}|^2, \nonumber \\
    f_{H} &= \sum_{j} \chi H^2 |\eta_0^j|^2, \label{eq:GLFE}
\end{align}
where $a_\mu=a^0_\mu (T-T_{{\rm c},\mu})$, with $a^0_\mu$, $b_\mu $ and $K_\mu$ positive, real phenomenological constants, and $\epsilon$ quantifies the effect of ASOC leading to the coupling between the even- and odd-parity OP and transforms under the $A_{2u}$ representation.
Moreover, $\vec{D}_\parallel=[-i \boldsymbol{\nabla}+2e \vec{A}]_\parallel$ is the in-plane covariant derivative.
Despite considering the situation in a magnetic field, we do not include the property of the mixed phase and, therefore, omit the vector potential $ \vec{A} $ when discussing the phase diagram.
The effect of the magnetic field then enters only through the Zeeman coupling to the electron spin, which involves paramagnetic limiting for the even-parity OP with $ \sum_j \chi | \eta_0^j|^2 $ yielding a reduction of the Pauli spin susceptibility.
While we introduced the bare critical temperatures for both OP components, their ASOC-induced coupling yields a higher effective onset of superconductivity, $ T_{\rm c} >  T_{{\rm c},\mu}$.

\begin{figure}[t]
    \centering
    \includegraphics[width=8.3cm]{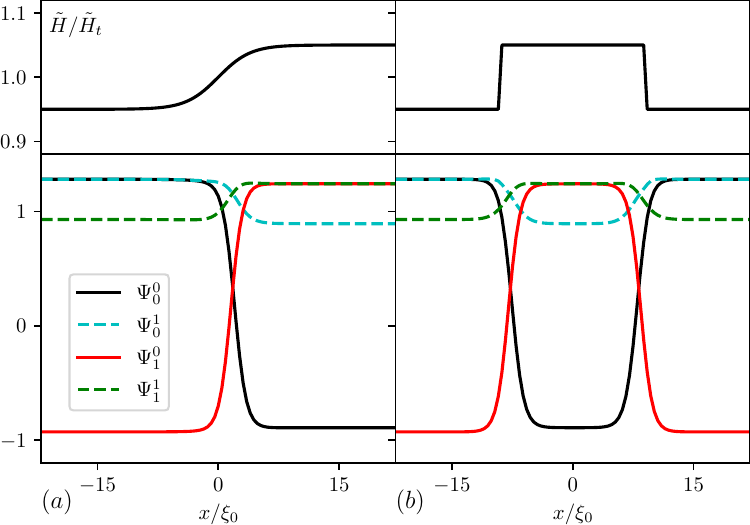}
    \caption{
    Effective magnetic field profiles (top row) and the resulting OP configurations (bottom row) computed via numerical minimization.
    (a) features hyperbolic tangent inhomogeneity with width $4\xi_0$, whereas (b) shows a double DW configuration.
    All $\tilde{H}(x)$ profiles are symmetric with respect to $\tilde{H}_t$ with a maximum deviation of $5 \%$. For the numerical calculations, we fix temperature to  $T/T_{c,0}=0.5$ and use
    the parametrization: $b_\mu=b$, $K_\mu=K$, $-a_0/b=K/|a_0|=1$, $a_1/a_0=0.105$, $J/|a_0|=1$, and $\epsilon/|a_0|=2$. Lengths are expressed in units of $\xi_0=\sqrt{K/|a_0|}$ and the OP as $\Psi_\mu^j=\sqrt{-b/a_0} \eta_\mu^j$.}
    \label{fig:DomainWalls}
\end{figure}

In the limit of vanishing interlayer coupling incorporated by the Josephson-like $J$-term, the staggered system consists of independent noncentrosymmetric layers, naturally hosting mixed-parity pairing states.
Due to the alternating sign of the $\epsilon$-term, the relative sign between the two OP components alternates as well, being ``$+$'' for even $ j$ and ``$-$'' for odd $j$ in our setup.
A finite $J$ explicitly restores inversion symmetry, enabling us to label the solutions as even and odd under inversion operation, which exchanges the two layers.
Assuming that at zero magnetic field the even-parity OP is dominant ($ T_{{\rm c},0} > T_{{\rm c},1} $), the interlayer coupling stabilizes a configuration, where $ \eta_0^j $ has the same phase in all layers while $ \eta_1^j $ alternates its sign.
This corresponds to the low-field A phase: $\eta_0^j=\eta_0^{j'}=u_A$, $-\eta_1^j=\eta_1^{j'}=v_A$ with $u_A > v_A > 0$ and $ j=2n $ (even) and $j'=2n+1 $ (odd). 
In contrast, in the high-field B phase, $ \eta_0^j $ is suppressed through the paramagnetic limiting effect such that the now dominant $ \eta_1^j $ has the same phase for all layers with an alternating sign for $ \eta_0^j $: $-\eta_0^j=\eta_0^{j'}=u_B$, $\eta_1^j=\eta_1^{j'}=v_B$ with $v_B > u_B > 0$, as described, for instance, in Ref.~\cite{schertenleib2021unusual}.
It is straightforward to show that the transition occurs at a magnetic field given by $H_t=\sqrt{(a_1-a_0)/\chi}$.
The model outlined in Eq.~\eqref{eq:GLFE} is geared towards capturing the internal phase transition and does not account for e.g. orbital depairing that ultimately destroys the B phase at high fields.
Nevertheless, the above formula for $H_t$ provides a good approximation even in the presence of the vortex lattice~\cite{schertenleib2021unusual}.
Because the inversion operation exchanges even and odd layers, we may characterize A as {\it even}- and B as {\it odd}-parity phase.

At sufficiently low temperature, both phases correspond to separate minima of the GL free energy, where at $ H = H_t $ they exchange their role as the global minimum and give rise to a first-order transition,
as can be shown also based on general thermodynamic arguments~\cite{leggett1974implications,yip1991thermodynamic}.
Importantly, in a spatially inhomogeneous material, we expect the transition to not happen everywhere at the same field $H_t$, but to show regions of one phase embedded within the other, whereby the OPs rearrange between even and odd configurations on a length scale of order coherence length $\xi_0=\sqrt{K/|a_0|}$, constituting sharp DWs.


\emph{Domain walls.}
Inhomogeneities in the sample lead to spatially varying parameters and thus, critical field. We incorporate such inhomogeneities in our GL model through a position-dependent parameter $\chi(\vec{r})$ assuming a relatively slow variation in space, which only enters the Zeeman term $f_{H}$ in the free energy.
In order to investigate the structure of a DW, we restrict ourselves to a one-dimensional spatial dependence, $ \chi(x) $, use the parametrization  $\sqrt{\chi(x)}H\equiv \tilde{H}(x)$, and, for a model calculation, consider a step-like dependence,  $ \tilde{H}(x) = H_0 + H_1 \tanh(x / \delta ) $ with $ H_0 - H_1 < \tilde{H}_t < H_0 + H_1 $ and $ \tilde{H}_t=\sqrt{a_1-a_0}$ corresponding to the transition point.
In this way, the A (B) phase is favored for $ x < 0 $ ($ x > 0 $).
We minimize the free energy numerically using the one-step relaxed Newton-Jacobi method~\cite{newton-jacobi} and obtain a smooth, narrow DW between the two phases, as displayed in Fig.~\ref{fig:DomainWalls}~(a).
Using a profile for $ \tilde{H}(x) $ with two opposite steps shows an ``island'' of B phase inside the A phase [Fig.~\ref{fig:DomainWalls}~(b)], as it may occur at the verge of the bulk phase transition. 
The position of the DW is pinned by the specific $x$ dependence of $ \tilde{H}(x) $, roughly at the position $ \tilde{H}(x_0) = \tilde{H}_t $, say $ x_0 = 0 $ in Fig.~\ref{fig:DomainWalls} (a). 
In what follows, we consider a harmonic pinning potential for the DW position.

Note that, unlike deep in the A or B phase, within the DW the OP does not adhere to global parity classification.
We also remark that the deformation of OPs in the vicinity of the DW combined with the local non-centrosymmetricity yields a non-vanishing spin current along the domain boundary~\cite{holst2022role,supplementary}.
However, we do not dwell on this feature in the following, but rather consider the effects of strain in the system.


\emph{Ultrasound probe.}
Recent experiments observed a significant effect of hydrostatic pressure on the superconducting phase diagram of CeRh$_2$As$_2$.
Most glaringly, pressure decreases the overall $T_{\rm c}$ and affects the balance between A and B phases~\cite{siddiquee2022tuning}.
Owing to its tetragonal symmetry, the phase diagram of CeRh$_2$As$_2$ under hydrostatic pressure can also be appreciated qualitatively by considering that strain along the $c$ axis changes the ratio between the strength of parity mixing and interlayer coupling, namely $\epsilon/J$. While this aspect may be more purposefully investigated by uniaxial $c$-axis strain, here we propose a related experiment based on ultrasound modes. 

For this purpose, we first discuss the coupling of DWs to ultrasound.
DWs represent a boundary between two phases with vortex lattices of different flux density and, thus, different mean magnetization~\cite{schertenleib2021unusual}.
Naturally, this induces a supercurrent flowing within the layers along the domain boundary.
The temporal variation of strain through the ultrasound waves leads to a periodic displacement of the DW and yields a dissipation via the motion of currents as well as the vortices near the DW.
Thus, we expect an extra contribution to ultrasound absorption due to local nucleation in the vicinity of the first-order phase boundary.

To formulate this concretely, we introduce an ultrasound mode as a longitudinal plane wave of frequency $ \omega$ propagating in the $z$ direction (crystallographic $c$ axis), $\vec{u}(\vec{r},t)= \hat{z} u_0 e^{i kz-i\omega t}$, where $ \vec{u} $ is the lattice displacement field and $ k $ is the wave vector along $z$, which depends on the properties of the medium.
This mode transforms under the trivial $A_{1g}$ representation and hence is not symmetry breaking.
Since the only relevant element in the strain tensor is $ \epsilon_{zz} = \partial_z u_z $, we focus in the discussion of the deformation energy and strain-OP coupling on this component only. 
Then the corresponding additional terms to the free energy include the elastic energy $f_{\rm el}$ and the coupling term, $f_{\rm s\eta}$, yielding~\cite{sigrist2002ehrenfest}
\begin{align}
   f_{\rm el} &= \frac{c_{33}}{2} \epsilon_{zz}^2 = \frac{c_{33}}{2} (\partial_z u_z)^2, \nonumber \\
   f_{\rm s \eta}  &= \sum_j \Big( \gamma_0 \epsilon_{zz} |\eta_0^j|^2 + \gamma_1 \epsilon_{zz} |\eta_1^j|^2\Big),
\end{align}
with $ c_{33} $ the elastic constant and $ \gamma_{0,1} $ phenomenological coupling constants.
These terms would be sufficient to describe the modification of the phase diagram under $c$-axis uniaxial strain.
The DWs are pinned by the spatial inhomogeneities of $ \tilde{H}(\vec{r} ) $, which we model by a harmonic potential around a given equilibrium position.
Strain then imposes a force to shift this position.
With this in mind, we write the DW potential term as
\begin{align}
       f_{\rm DW} &= \frac{1}{2}c(\tilde{H}) \Bar{x}^2 + \gamma \epsilon_{zz} \bar{x},
\end{align}
where $\bar{x}$ is the position of the DW, which rests at $ x =0 $ in the absence of strain.
For simplicity, we assume a DW that is infinitely extended in $y$- and $z$-direction such that we can model our system in one spatial dimension, $x$. 
The coefficient of the quadratic term $c(\tilde{H})$ depends on the details of $\tilde{H}(x)$, and by linearizing we find to leading order that $c(\tilde{H})\propto (\partial \tilde{H}(x)/\partial x )\rvert_{x=0}$~\cite{supplementary}.
Intuitively, a steeper inhomogeneity profile yields a stiffer potential restricting the DW motion. 
On the other hand, $\gamma$ is independent of $\tilde{H}(x)$ and can be approximated through the different strain dependence of energy densities in the A and B phases, $f_A$ and $f_B$ respectively, such that~\cite{supplementary}
\begin{align}
    \gamma\propto\bigg(\frac{\partial f_A}{\partial \epsilon_{zz}}-\frac{\partial f_B}{\partial \epsilon_{zz}}\bigg) \bigg\rvert_{\epsilon_{zz}=0}. \label{eq:gamma}
\end{align}
Note that the equilibrium position $\Bar{x}=-\gamma\epsilon_{zz} /c(\tilde{H})$ still depends on the details of $\tilde{H}(x)$.

We model non-resonant absorption by the overdamped oscillation of the DW position and write the equation of motion for $\bar{x}$ as well as the displacement field $u_z$ as
\begin{align}
   \eta \partial_t \bar{x}=-\frac{\partial f_{\rm tot}}{\partial \Bar{x}},
     \hspace{0.4cm}
   \rho \partial_t^2 u_z(\vec{r},t) = \sum_i \partial_i \frac{\partial f_{\rm tot}}{\partial[ \partial_i u_z(\vec{r},t)]}.
\end{align}
Here $f_{\rm tot}= f+f_{\rm el}+f_{\rm s \eta}+f_{\rm DW}$, $\eta$ is a phenomenological viscosity constant, resulting from the dissipation due to the motion of the DW, and $\rho$ is a mass density. 
For an external drive of frequency $\omega$, the resulting wave vector reads
\begin{align}
    k=\frac{\sqrt{\rho}\omega}{\sqrt{ c_{33}- \frac{\gamma^2}{ i\omega \eta + c} }}.
\end{align}
This finally yields the ultrasound absorption coefficient~\cite{supplementary}
\begin{align}
    \alpha&=-{\rm Im}\ k= \frac{\gamma^2}{2 c_{33} c_s \eta} \frac{\omega^2}{\omega^2 + \omega_0^2},\label{eq:alpha}
\end{align}
where $ c_s = \sqrt{c_{33}/\rho} $ denotes the sound velocity and $\omega_0=c/\eta$ is a characteristic frequency of the DW. The coefficient $\alpha$ is a measure for the absorption per unit area of the DW.
The total absorption is proportional to the spatial average of $\alpha$, as well as the DW density in the system for a given external magnetic field. Below we elaborate on the evolution of the latter as one tunes the magnetic field across $H_t$.
As pointed out earlier, larger value of $c$ (steeper inhomogeneity profile) describes stronger pinning, which in turn results in less movement of the DW and therefore less absorption, which is reflected in Eq.~(\ref{eq:alpha}).
Hence, $c$ is in general position dependent.
The origin of the viscosity parameter $\eta$ is analogous to that of vortex motion in the mixed phase of a superconductor, but depends on the microscopic details of the DW.

Our simple model accounts for the contribution to ultrasound absorption per unit area of a DW. As such, the total ultrasound absorption depends on the DW density upon driving the external magnetic field through the phase boundary at $H_t $.
As the magnetic field approaches $ H_t $ from below, islands of the high-field phase B appear due to inhomogeneities within the A phase, as schematically shown in Fig.~\ref{fig:phasediag}(b). In our one-dimensional model, such a B-phase bubble is depicted in Fig.~\ref{fig:DomainWalls}(b).
Upon increasing the field, the islands bounded by a DW proliferate and eventually DWs percolate through the sample.
A further rise of the magnetic field leads to the shrinkage of A domains in a landscape dominated by the B phase [see Fig.~\ref{fig:phasediag}(b)].
Naturally, the largest contribution to sound absorption is expected when the DWs percolate and their density is largest, such that the absorption shows a peak as a function of the external field.

Finally, let us make some practical remarks on the experimental signatures that would help identify the absorption mechnanism in a material like CeRh$_2$As$_2$. 
To date, the first-order nature of the internal phase transition has not been unambiguously established. 
Since a possible second-order phase transition could also yield an absorption peak upon tuning across the transition line (analogous to critical opalescence), the observation of an absorption peak alone may not be conclusive.
However, if temperature is lowered when the system is on the verge of a second-order phase transition, the critical region narrows down with the accompanying absorption peak becoming sharper.
On the other hand, the superconducting coherence length, given in our model by $\xi_0=\sqrt{K/|a_0|}$, decreases with decreasing temperature, whereby the OPs can realize a DW on a smaller length scale, thus in a wider magnetic field range. 
For this reason, a first-order phase transition with accompanying local nucleation tends to broaden the absorption peak as function of field when the temperature is lowered, as the opposite phase can nucleate more easily on smaller inhomogeneities.
The evolution of an absorption peak as a function of temperature may therefore help pin down the order of the internal phase transition in this system.


\emph{Discussion.}
While the proposed mechanism of ultrasound absorption provides a way to detect the first-order transition between the phases A and B in the presence of sample inhomogeneity or inhomogeneous magnetic field, the non-uniform nucleation could also be observed by other means.
As mentioned earlier, the vortex-lattice density is different in the two phases~\cite{schertenleib2021unusual} and it may be possible to track this inhomogeneity by surface scanning probes either by observing the magnetic field or the vortex lattice on its own.
A further interesting aspect is provided by the different spin susceptibilities of the A and B phases.
Nuclear magnetic resonance (NMR) Knight-shift measurements as a local probe would in principle be able to observe two distinct signals with a varying volume fraction of the two domains as the magnetic field is tuned through $ H \sim H_t $. 


This work was motivated by the heavy fermion superconductor CeRh$_2$As$_2$.
Some aspects of our discussion, however, might not apply straightforwardly to this material.
Recent NMR measurements indicate the presence of magnetic order coexisting with the A phase, but not with the B phase~\cite{kibune2022observation,ogata2023parity}.
We did not attempt here to integrate this feature into our discussion, since it is unclear so far how this material-specific property connects to the field-induced (first-order) transition. 
An interesting point we also leave open for the time being is the study of damping mechanism.
It will be interesting to see how the DW shape and the supercurrents influence the viscous motion of DWs. Naturally, such questions will become more urgent once the presence of such DWs is established. 
Still, besides probing the physics of nucleation in a novel, superconducting context, our efforts may help further understand the phase diagram of CeRh$_2$As$_2$, where investigations are still in their early stages.

\emph{Acknowledgement.}
We thank Elena Hassinger, Manuel Brando, Stanislaw Galeski, and Maximilian Holst for useful discussions. A.S. and M.S. are grateful for financial support from the Swiss National Science Foundation (SNSF) through Division II (No. 184739). M.H.F. also acknowledges financial support from SNSF through Division II (No. 207908).
\bibliography{Library_CeRh2As2}

\end{document}